# Important notice

This preprint has not undergone peer review or any post-submission improvements or corrections. The Version of Record of this article is published in Nature Geoscience, and is available online at http://doi.org/10.1038/s41561-026-02089-9 under the title "**Homogeneous ice Nucleation from Water Vapour Suggested by Elongated Clouds on Mars**", scheduled for publication in October 7, 2026.

According to the publication agreement, the accepted version of the manuscript (highly improved thanks to the voluntary work of three expert reviewers) cannot be published in open repositories until 6 months (Embargo period) from the publication date in Nature Geoscience. This is because we chose not to pay the ~12 500€ of fees for open access publication in Nature Geoscience.

Note then that the text of **this document is not the definitive text of the article, and it contains known errors and inaccuracies that were corrected during the review process. The final text is much better and includes new analysis, thanks to external reviewer's comments, but for now we cannot make that text freely available.**

# Homogeneous Nucleation of Water Vapor Evidenced by Elongated Clouds on Mars

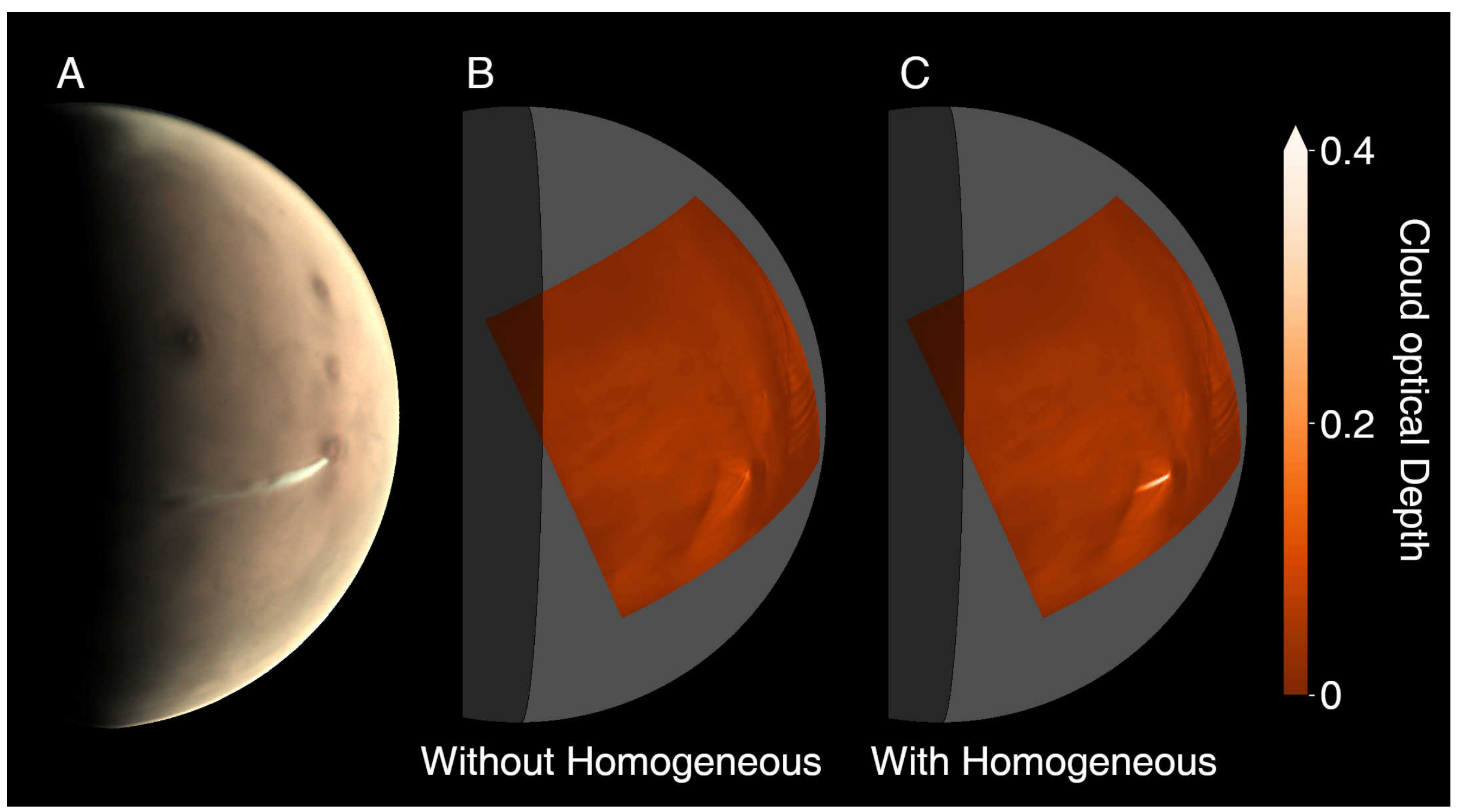

# Homogeneous Nucleation of Water Vapor Evidenced by Elongated Clouds on Mars

J. Hernández-Bernal[1], A. Määttänen[2], A. Spiga[1], F. Forget[1]

[1]Sorbonne Université, ENS Paris, Université PSL, CNRS, École polytechnique, Institut polytechnique de Paris, Laboratoire de Météorologie Dynamique, LMD, Place Jussieu 4, 75005, Paris, France

[2]LATMOS/IPSL, Sorbonne Université, UVSQ Université Paris-Saclay, CNRS, Paris, France

## Abstract

It is widely accepted that water vapor forms clouds by heterogeneous nucleation (condensation on pre-existing dry nuclei). Homogeneous nucleation (directly from water vapor, without a pre-existing nuclei) is theoretically possible, but it requires high levels of supersaturation and so it is not considered a viable cloud formation process under real atmospheric conditions. Here, we show that introducing homogeneous nucleation in a meteorological model of Mars allows to successfully reproduce the unique characteristics of the Arsia Mons Elongated Cloud, a remarkable water ice cloud that occurs recurrently on Mars and which had remained impossible to model under conventional cloud microphysics. This finding demonstrates that homogeneous nucleation happens on Mars, provides the first evidence of homogeneous nucleation of water vapor in a planetary atmosphere, and challenges current assumptions about cloud formation processes, which is also relevant for certain clouds on Earth and potentially on other planets.

## Main text

The formation of cloud particles from water vapor can theoretically happen either by heterogeneous or homogeneous nucleation[1-3]. Heterogeneous nucleation occurs when water vapor condenses onto preexisting aerosol particles. Homogeneous nucleation consists on particles forming purely from vapor, without a preexisting solid surface. While both processes are theoretically possible, homogeneous nucleation of water vapor requires very high saturation ratios (relative humidity) compared to heterogeneous nucleation. Consequently, heterogeneous nucleation is expected to dominate water cloud formation in planetary atmospheres, where aerosols are typically present, unless supersaturation reaches exceptional levels[1-3].

In the Earth's atmosphere condensation of water vapor can happen from vapor to liquid, in the troposphere; or from vapor to ice, in the mesosphere; depending on the atmospheric vapor pressure with respect to the triple point of water. In both cases, homogeneous nucleation is considered impossible under real atmospheric conditions, in which only heterogeneous nucleation is considered[1-3]. The occurrence of homogeneous nucleation in the mesosphere has been sporadically discussed as a theoretical possibility[4-5]. Yet, no evidence of it has been reported, and it is usually regarded improbable[6-10]. As a matter of fact, research on cloud formation processes has focused on heterogeneous nucleation[11-14].

On Mars, clouds form well below the triple point of water, and thus water vapor condenses into ice. Cloud formation has also been considered to occur exclusively via heterogeneous nucleation[15-18]. Early analysis showed that homogeneous water ice nucleation on Mars would require[16] a saturation ratio above ice of the order of $10^5$. This saturation ratio was deemed implausible, because the abundant dust is expected to deplete water vapor in excess of saturation via heterogeneous nucleation before such extreme supersaturation could develop. Nevertheless, both observations[19-21] and model simulations[22,23] have revealed that substantial levels of supersaturation can be found in the Martian atmosphere.

The Arsia Mons Elongated Cloud (AMEC; Fig. 1A) is a strikingly elongated, recurrent cloud that forms each morning on the Western slope of the Arsia Mons volcano at approximately 45 km altitude[24]. It occurs during a part of the dusty season of Mars, when the atmosphere is dust loaded and warmer, displacing the hygropause upwards and making clouds less common in low latitudes. Every day, the elongated cloud starts to form at sunrise, undergoes rapid horizontal expansion over 3 hours, and subsequently detaches from the volcano. At its peak, it can stretch up to 1800 km, growing through wind-driven advection, as revealed by time-lapse imaging[24]. Comparable, though smaller, clouds have also been observed during the same season and at similar altitudes in nearby regions[25,26].

Various processes have been suggested to explain the distinctive morphological and dynamical characteristics of these elongated clouds, including the vertical injection of water by slope winds[24], and cloud formation within cold pockets generated by vertically propagating gravity waves, which is supported by model simulations[23,26]. Yet none of those proposed mechanisms has hitherto convincingly reproduced the elongated tails of these clouds.

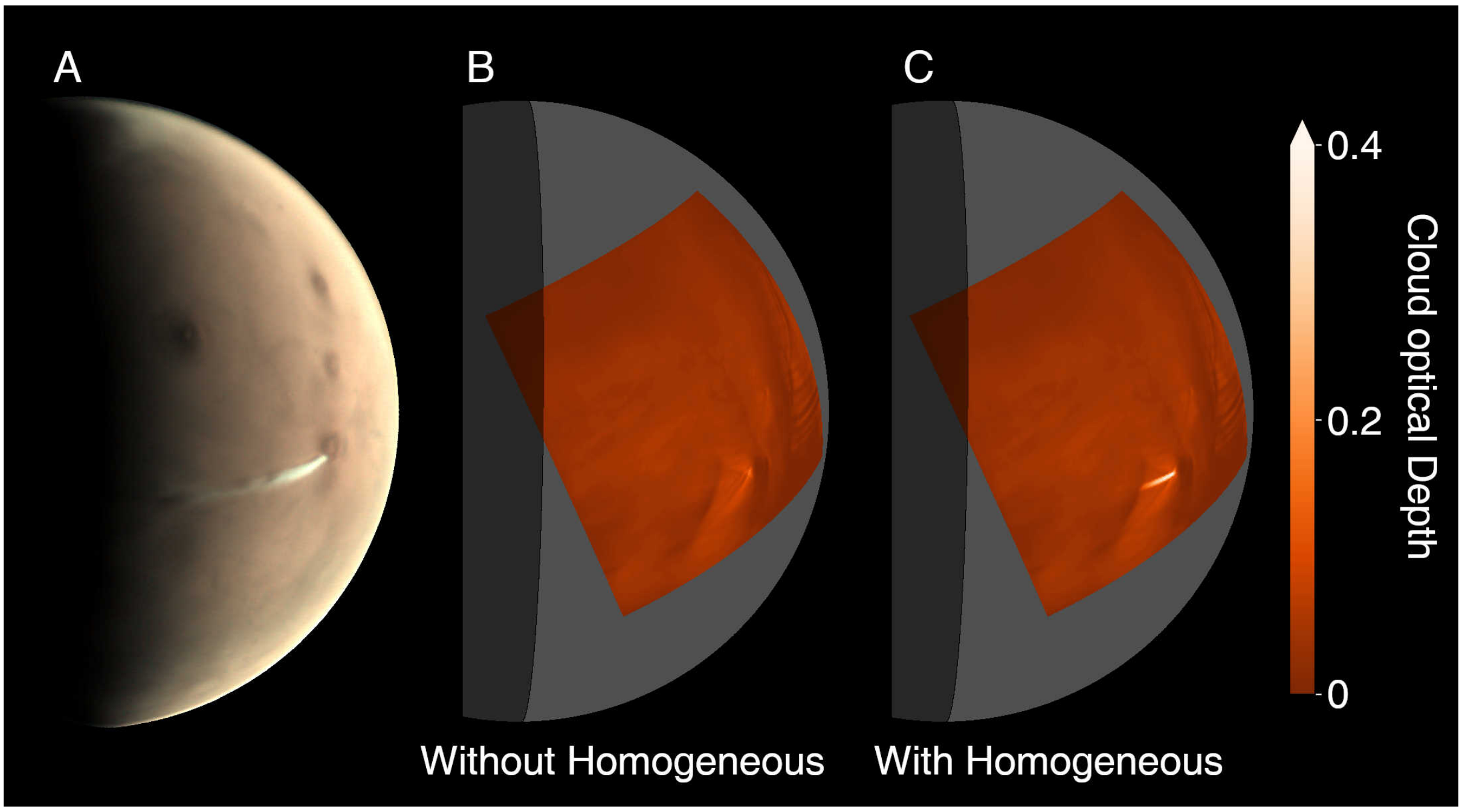

**Figure 1. Observation and modeling of the Arsia Mons Elongated Cloud (AMEC).** (A) Observation by the Mars Express VMC imager[41,42], at 8.5 a.m. Local Time. (B) Reference simulation assuming only heterogeneous nucleation, at 8.1 a.m. Local Time. (C) Same as B, but with homogeneous nucleation added. See also supplementary video S1.

We investigate the Arsia Mons Elongated Cloud using a Mars mesoscale model[27,28,23] consisting of the hydrodynamical solver of the Weather Research and Forecast model[29] coupled with the physical packages developed in the framework of the Mars Planetary Climate Model[30](PCM; see Methods).

In accordance with prior studies[23], simulations with this mesoscale model employing standard cloud microphysics assuming heterogeneous nucleation on mineral dust particles[22,31] fail to reproduce the observed elongated cloud tail (Fig. 1B). Nevertheless the model captures the orographic generation of vertically propagating quasi-stationary gravity waves (streamlines in Fig. 2A), that induce strong adiabatic cooling over a broad region West of Arsia Mons, with temperatures lower than the environment average (~160K at 45 km altitude), and supersaturated water vapor (Fig. 2B). A compact volume where temperatures are the lowest – "core cold pocket" from now onwards – appears on the Western slope of Arsia Mons (Figs. 2 A,B), in the same position where the cloud origin is observed. The maximum saturation ratio reached in the core cold pocket is of the order of $10^5$.

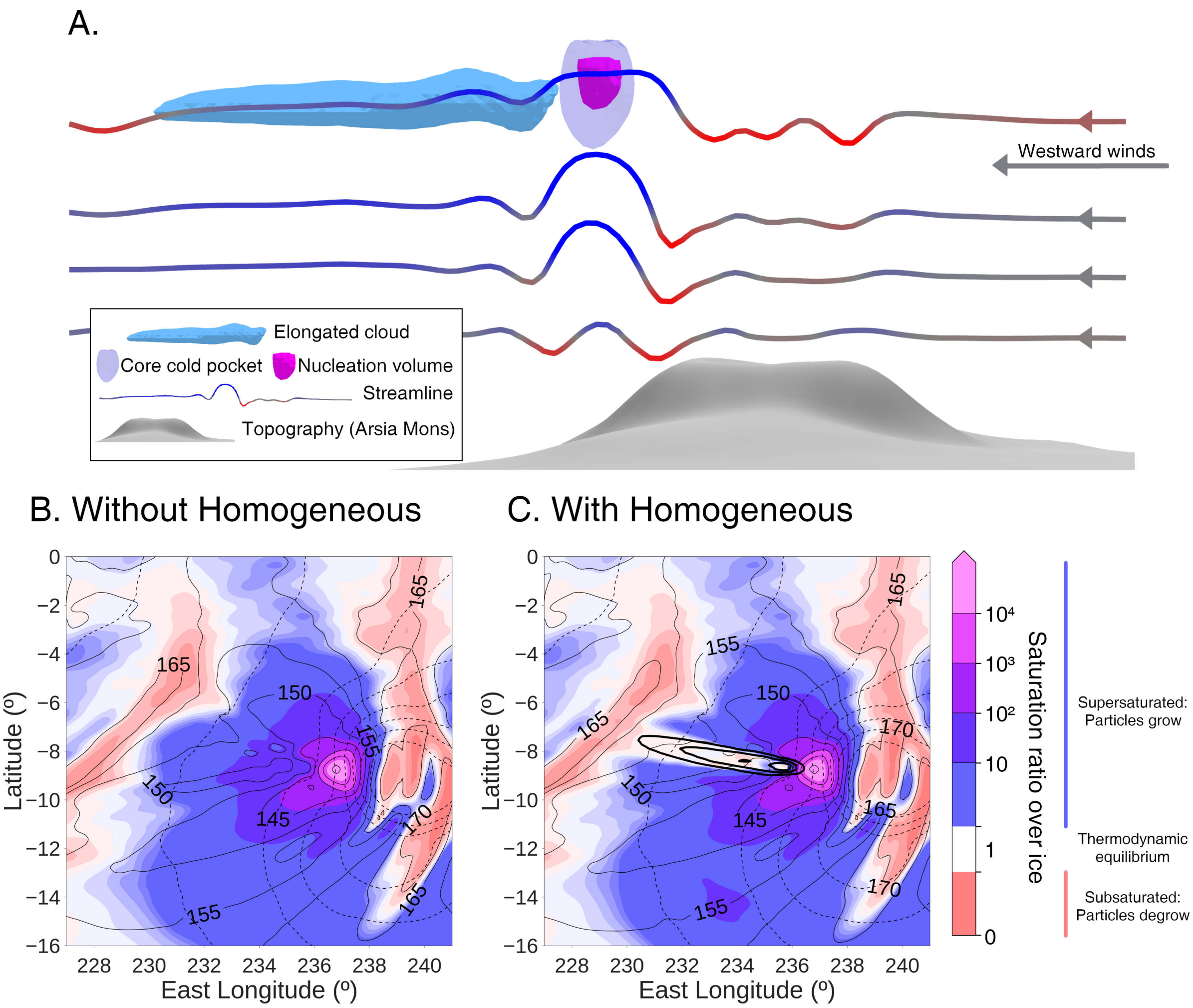

**Figure 2. Key processes involved in the Arsia Mons Elongated Cloud**. From model outputs at 8.1 a.m. Local Time (A) Schematic representation of cloud formation, directly from model outputs. Streamlines represent orographically generated quasi-stationary gravity waves; red/blue coloring indicates adiabatic warming/cooling respectively. (B) Environment saturation (background) in the reference simulation with only heterogeneous nucleation at 45 km of altitude. The temperature field (K) is represented by thin numbered contours. Dashed contours represent topography. (C) Same as B, but for the reference simulation with homogeneous nucleation. Thick contour lines represent the homogeneously formed elongated cloud.

Air parcels approaching the core cold pocket are displaced upwards by several kilometers, experiencing a considerable adiabatic cooling rate of 0.05 K/s sustained over 800 seconds (Extended Data Fig. 1), and reaching a final temperature below 130K. The modeled atmosphere in this region contains ~70ppmv of water vapor and ~0.1 $cm^{-3}$ dust particles (Extended Data Fig. 2), which is consistent with remote-sensing observations of these variables at this area and season[32-34].

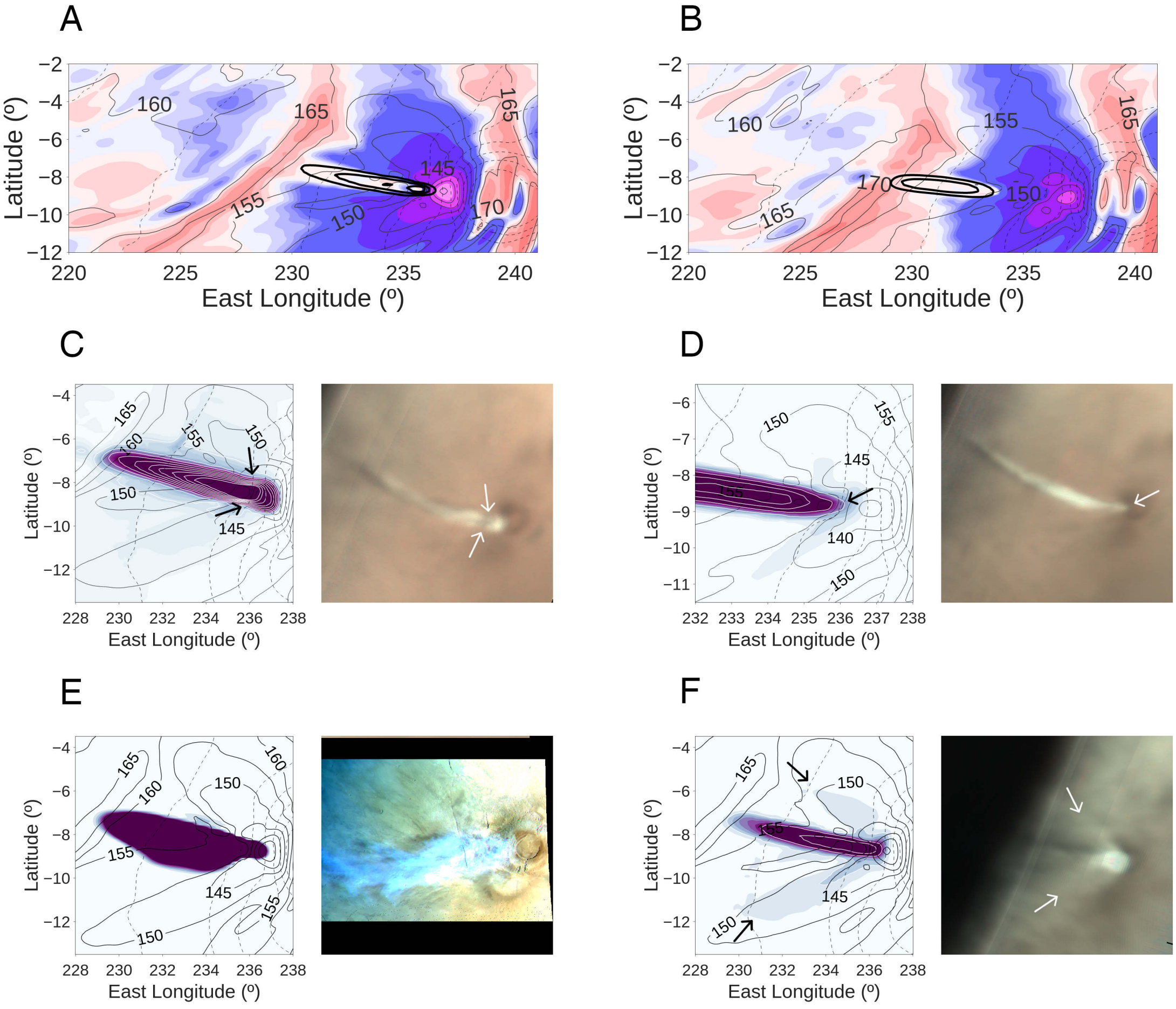

**Figure 3. Detailed comparison of observed and modeled features of the Arsia Mons Elongated Cloud.** Thin black contours represent temperatures at 45 km, dashed contours represent topography. (A,B) Same as 2C. with (A) before detachment at 8.1 a.m. Local Time. And (B) after detachment at 8.9 a.m. Local Time. (C-F) Each panel displays the cloud optical depth from the mesoscale model in the left and a real image for comparison in the right. Visualization level of optical depth were adjusted to highlight specific details in each case C(0.05 to 0.4), D(0.1 to 0.3), E(0.05 to 0.4), F(0.08 to 0.2). (C) Narrow neck, around 7.5 a.m. Local Time. (D) Narrowing before detachment, around 8.5 a.m. Local Time. (E) Outbreak reproduced when deactivating heterogeneous nucleation, at 8.1 a.m. Local Time. (F) V-shaped hazes, at 8.1 a.m. Local Time in the reference simulation with homogeneous nucleation compared to observations around 6.0 a.m. Local Time.

We analyzed the microphysics of air parcels traversing the core cold pocket, with the aid of a 0D microphysics model (see Methods). This analysis reveals that hundreds of ice particles per cubic centimeter nucleate homogeneously under these extreme conditions, with a peak nucleation rate of ~4 $cm^{-3}s^{-1}$ (Extended Data Figs. 1,3).

We incorporated homogeneous ice nucleation into the mesoscale model, and found that model could then successfully reproduce the AMEC (Fig. 1C). The extreme conditions present in the core cold pocket trigger rapid formation of a dense population of homogeneous ice crystals (Extended Data Fig. 4), which are then transported by westward winds (Fig. 2A). Those cloud particles grow until they reach thermodynamic equilibrium with surrounding air (saturation ratio = 1; Fig. 2C). In contrast, other areas of the atmosphere are supersaturated (saturation ratio >1), because dust concentrations and ice particle growth rates are too low to deplete water vapor in excess of saturation and produce a significant cloud via heterogeneous nucleation. This is the physical basis of the extraordinary length of the cloud tail: it is in thermodynamic equilibrium, and so once formed, it remains stable as it is advected away from the core cold pocket.

Later in the day, when the core cold pocket fades, homogeneous nucleation ceases, and the cloud tail detaches from its origin in simulations (fig. 3A-B; supplementary videos S1 and S2), as it does in observations. Prior to this detachment, the head of the cloud has been observed to shrink in size (Fig 3D), a behavior reproduced by the model through the progressive reduction of the area where homogeneous nucleation occurs, as the core cold pocket fades.

A narrowing of the tail right after the head of the cloud, and variations in brightness along the tail are sometimes observed (fig. 3C), and similar features are found in our simulations. They are explained by variations in the thermal field traversed by the cloud particles, which lead to their transient growth and degrowth.

The rarely observed “outbreak” of the cloud (fig. 3E) happens in our model when we deactivate heterogeneous nucleation, showing how the presence of dust moderates the amount of homogeneous nucleation via competition with heterogeneous nucleation. Heterogeneous nucleation also plays a role in reproducing certain observations, such as the occasionally observed V-shaped hazes, that are formed in the model by heterogeneous nucleation in certain areas of the thermal field (fig. 3F).

To sum up, the different defining features of the observed AMEC are qualitatively reproduced in the mesoscale model only once homogeneous nucleation is considered. The simulated optical depth (~0.4) and the cloud particle size (~0.25 μm) in the reference simulation with homogeneous simulation are consistent with observations. There is room for improvements in other quantitative parameters: the modeled cloud is 30-50% narrower than in observations, its onset is delayed by ~1.5 hours, and its maximum length prior to detachment is ~25% of that observed, which is likely explained by the delay in the starting, the slower winds, and the presence of a warm region in which part of the cloud particles sublimate. Details on this quantitative comparison are given in Supplementary Discussion S1.

We conducted additional mesoscale simulations to assess the sensitivity of the model to poorly contrained physical parameters and to the background atmosphere (see Methods and Extended Data Table 1). Modeled cloud characteristics such as width, length, onset time, and optical depth were found to be sensitive to the amount of homogeneous nucleation (Extended Data Figure 5). This reveals that observed day-to-day variations in these parameters[24] emerge as the result of fluctuations in meteorological variables that control homogeneous nucleation.

Mesoscale simulations with homogeneous nucleation successfully reproduce observations of the elongated cloud, while simulations with only heterogeneous nucleation on regular micron-sized dust particles do not. However, beyond this regular coarse dust, a secondary Aitken mode (finer dust, with particle sizes<100nm) might be present in the atmosphere of Mars[35-37]. Could heterogeneous nucleation on this Aitken dust prevent homogeneous nucleation and provide an alternative pathway to reproduce observations?

Using the 0D microphysics model, we calculated that homogeneous nucleation still produces an optically thick cloud downwind from the core cold pocket assuming Aitken dust number densities of up to 1-20 $cm^{-3}$ (depending on the contact parameter; see Extended Data Figure 3).

We also introduced the Aitken mode in the mesoscale model (see Methods and Extended Data Table 1), and we did not succeed to produce the observed AMEC tail by heterogeneous nucleation on Aitken dust. These simulations show that microphysics on Aitken dust tend to produce widespread hazes (Extended Data Figure 6) which in some cases resemble cloud patterns observed when the atmosphere is very dusty, but never the elongated cloud (Extended Data Figure 7). In some simulations the hazes are thicker downwind from the core cold pocket, mimicking the elongated cloud (Extended Data Figure 6), but remains diffuse and not optically thicker than surrounding hazes. In contrast, observations of the AMEC with the Mars Express HRSC imager[38,39] display a very well contrasted cloud tail in terms of optical depth[40] (Extended Data Figure 8). Analysis with the 0D microphysics model shows that a meridional gradient in the thermal field much larger than predicted by the mesoscale model would be required to produce the level of optical depth contrast observed in the AMEC through heterogeneous nucleation on Aitken dust (Extended Data Figure 9). This analysis sets an upper limit for the Aitken mode during the occurrence of the AMEC (1-20 $cm^{-3}$; Extended Data Figure 3), that cannot be verified with observations due to the current lack of data of the Aitken mode during this season.

Our investigations demonstrate that the spectacularly-elongated water-ice cloud recurrently observed on Mars above the Arsia Mons volcano (the aforementioned AMEC) is the result of homogeneous nucleation happening in the Martian atmosphere. This conclusion frames the AMEC as an outstanding manifestation of homogeneous nucleation directly from water vapor in a planetary atmosphere, a phenomenon taught but considered theoretical, never observed in nature before.

In the case of the AMEC, homogeneous nucleation is enabled by a very dynamical situation in which moist air parcels are lifted by several kilometers in only a few minutes, producing a large and rapid adiabatic cooling that favors homogeneous nucleation over heterogeneous nucleation, the latter being limited by the low abundance of dust. We will explore in future works the extent and role of homogeneous nucleation on Mars. Is homogeneous nucleation a rare phenomenon that only takes place in the AMEC, or does it happen in other cloud formations? Does it play any relevant role in the global water cycle? Detailed studies based on high resolution models and a careful interpretation of observations will be required to shed light on these questions.

The finding that homogeneous nucleation of water vapor can have significant effects in the atmosphere of Mars, calls for further research on the occurrence of this process elsewhere and in particular in the terrestrial mesosphere, where it is theoretically possible[4-5] but is often regarded as improbable or a mere theoretical curiosity in current literature[1-3,6-14]. Further research will be required to improve our understanding of homogeneous nucleation of water vapor, under low-temperature low-pressure conditions, present on both the terrestrial and the Martian atmosphere, and potentially on other planets.

# Methods

## Parametrizations for homogeneous ice nucleation from water vapor

Our modelization is based on the Classical Nucleation Theory as presented in Määttänen et al.[16]. Our modelization differs from that of Määttänen et al. in the parametrization of the surface energy (σ) and the vapor pressure of water with respect to ice, following updates in these parameters after the works of Murray and Jensen[5], and Nachbar et al.[44,45]. Also considering that according to these more recent works, homogeneous ice nucleation from water vapor happens as Amorphous Solid Water (ASW).

## Vapor pressure of ASW

Nachbar et al.[45] reported, based on laboratory experiments, the vapor pressure of ASW at temperatures between 140K and 160K, which are close to the temperatures relevant to this study.

The vapor pressure of ASW is:

$$P_{ASW} = P_{hex} \cdot \exp\left(\frac{\Delta G_{ASW \to hex}}{RT}\right) \quad \text{[Eq. 1]}$$

With:

$$\Delta G_{ASW \to hex} = (2312 \pm 227)\ J\ mol^{-1} - T\,[K] \cdot\ (1.6 \pm 1)\ \ J\ mol^{-1} K^{-1} \quad \text{[Eq. 2]}$$

And $P_{hex}$ is the vapor pressure of hexagonal ice, which is given by Murphy and Koop[46] as:

$$P_{hex}\ [Pa] = \exp\left(9.550426 - \frac{5723.265}{T[K]} + 3.53068 \cdot\ ln\,(T\ [K]) - 0.00728332 \cdot T\ [K]\right) \quad \text{[Eq. 3]}$$

By default, we use the vapor pressure for ASW derived by Nachbar et al. (Eqs. 1,2). However, in some sensitivity tests we use Eq. 3. See Extended Data table 1 for details.

## Surface energy of Amorphous Solid Water (ASW)

The homogeneous nucleation rate is highly sensitive to the surface energy of ASW, which is taken as the surface tension of liquid water[5]. Unfortunately, measurements of this parameter are only available[47] down to ~242K, and so there are large uncertainties in its value at temperatures relevant to us (120-140K). A report by the International Association for the Properties of Water and Steam (IAPWS) gave values for the surface tension of liquid water at temperatures over 0ºC, and proposed the following interpolation formula[48]:

$$\sigma\ [N/m] = 0.2358 \cdot \left(\frac{T_c - T}{T_c}\right)^{1.256} \left(1 - 0.625 \cdot \frac{T_c - T}{T_c}\right) \quad \text{[Eq. 4]}$$

With $T_c = 647.15K$.

More recently a new release by the IAPWS has confirmed[49] the validity of this formula "down to at least -25ºC". When used for extrapolation to low temperatures, this formula is well coincident with the results of a theoretical thermodynamic model[50], and so it has been employed as parametrization at temperatures around ~100K in some studies on homogeneous ice nucleation from water vapor[5,9].

### Parametrizations for heterogeneous ice nucleation: Temperature dependent contact parameter

The description of heterogeneous nucleation in our modelization is the one used in the Mars PCM models, described in various references[22,36,51].

Heterogeneous nucleation is sensitive to the particle size and to the contact parameter, which represents the ability of nucleation cores to nucleate water vapor. The contact parameter is nondimensional, and being defined as the cosine of the contact angle, it ranges from 1 to -1. Higher values of the contact parameter are associated to a higher probability of nucleation.

The contact parameter of Martian dust is not well constrained. Laboratory experiments with Martian dust analogues suggest high values around 0.95 at temperatures over 180K, but below 180K the contact parameter becomes temperature dependent. Määttänen & Douspis[52] have analyzed existing laboratory experiments and proposed various parametrizations for the temperature dependent contact parameter. The hyperbolic tangent parametrizations proposed by Määttänen & Douspis are in practice the most convenient to be implemented in models, because they are constrained to physically realistic values. In order to perform sensitivity tests, we employ two different parametrizations from Määttänen & Douspis[52]: The hyperbolic tangent based on experiments with the Mars analog sample JSC-1[53] (HTJSC), and the hyperbolic tangent based on experiments by Trainer et al.[54] (HTTRA).

Navarro et al.[22] used in the Mars PCM a constant value of 0.95 for the contact parameter, because it was the option that optimized the global water cycle. Recent improvements in the modeling of the water cycle have successfully introduced the temperature dependence of the contact parameter, and the PCM uses now the HTJSC parametrization instead[51]. Therefore, this is the parametrization for the contact parameter that we use here by default.

### Description of the Mars PCM and the mesoscale model

The Mars PCM is a GCM (Global Circulation Model) originally adapted to Mars by Forget et al.[30]. It employs a dynamical core from a terrestrial GCM coupled with dedicated models for sub-grid-scale physical processes occurring on Mars: radiative transfer, turbulence, microphysics[55,56,22]. These models of physical processes are continuously improved[57,58,51] and compared against observations [32,59,60,40].

The mesoscale model[27,28,23] employs the WRF (Weather Research and Forecast) dynamical core[29], and the same set physical packages as the PCM (different versions in some cases). The mesoscale model uses initial and boundary conditions from the GCM. The specific domain and methods employed here is the same as in Hernández-Bernal et al.[23] The domain covers the whole area of Tharsis, with 321 x 321 grid points and a spatial horizontal resolution of 10 km. In the vertical dimensions, 101 layers were simulated up to an altitude of 1Pa (approximately 60 km above the local surface). The first 24 simulated hours are considered as spin-up time and not considered in analysis. The dynamical timestep is 20s, and the WRF dynamical core uses the non-hydrostatic option. The physical timestep is 20s, except for microphysics, for which it is 10s. As in ref. *23*, simulations were run at Solar Longitude 270º with a dust scenario provided by Montabone et al.[61,62] corresponding to Martian Year 34.

The microphysical schemes employed in the Mars PCM physics were originally introduced and described by Navarro et al.[22]. They make use of two-moments log-normal distributions, meaning that particle sizes are assumed to follow a log-normal distribution described by two-moments (the number mixing ratio and the mass mixing ratio). This assumption enables a computationally efficient simulation of the particle sizes. The source code of modifications, including homogeneous nucleation of heterogeneous nucleation on Aitken dust, and the initial and boundary conditions employed in our mesoscale runs have been archived with the other datasets of this work[43].

## Inclusion of homogeneously formed clouds in the mesoscale model

Homogeneously formed ice crystals have some key differences with heterogeneously formed ones. Heterogeneously formed ice crystals have a dust core and that has implications for their treatment in the model. For a proper modelling of homogeneously formed crystals, we created a new independent scheme in the microphysics code, duplicating the one already existing for heterogeneous particles and making the necessary changes. We added two new tracers accounting for the number of homogeneously formed ice particles and the mass of ice in those particles, in analogy to the equivalent scheme employed for heterogeneously formed crystals[22]. The microphysics scheme for homogeneously formed crystals includes nucleation, growth, sublimation, and sedimentation. These new particles were not included in radiative transfer schemes, whereas the heterogeneously formed clouds are radiatively active in the model[22].

## Inclusion of heterogeneous nucleation on Aitken dust in the mesoscale model

We introduced a new scheme parallel to the one of heterogeneous and homogeneous cloud particles, introducing five tracers accounting for the number and mass of Aitken dust, the number and mass of nucleated Aitken particles, and the mass of ice condensed around nucleated Aitken particles. The Aitken dust is then modelled similarly as a prescribed two-moment log-normal distribution of particles. For simplicity, it is not currently coupled to the dust cycle nor to the radiative transfer. The Aitken dust does not sediment in the model (except through scavenging by cloud crystals) and it only interacts with the cloud microphysics. We can configure the initial effective radius and number density. This dust population in our model is only intended for making short-term and localized experiments regarding its interaction with microphysics; the proper investigation of Aitken dust in the atmosphere of Mars is left for future work. Aitken dust particles were simulated with an effective radius of 50nm, following the results by Fedorova et al.[35]

### Use of a 0D microphysics model

This model that we developed uses the same modeling of the fundamental microphysical processes (nucleation, particle growth) as the PCM models, but the implementation is different in the way it tracks particle sizes. It is a sectional model that discretizes the size distribution into bins to track particle sizes. In the case of heterogeneous cloud particles, it tracks simultaneously the size of the core and the ice shell. This approach is more sophisticated than the two-moment log-normal distributions employed in the PCM GCM and mesoscale models. In a 0D model, no vertical transport, such as sedimentation, is included.

The initial distribution of dust is set using log-normal distributions, with effective radius of 700nm for dust particles of the coarse mode, and 50nm for Aitken dust particles, following the results by Fedorova et al.[35] We use outputs from the mesoscale model to extract timeseries of temperature and pressure for a specific moving parcel, and we use these timeseries (with a smoothing scheme) in the 0D model. This implies that this model is equivalent to a Lagrangian air parcel moving in the mesoscale model. The results from the 0D model can be expected to differ slightly from those of the mesoscale model, due to the aforementioned key differences between the two models: the more sophisticated scheme to track particle sizes and numbers, and the Lagrangian approach.

### Relations between optical depth, particle number, particle size, and water budget

These four variables are interconnected and they are relevant for comparison to observations and budgets of water and numbers of particles:

- $\tau$: Cloud optical depth
- N: Particle number density (particles/volume)
- r: Radius of cloud particles
- $\chi_{H_2O}$: Volume mixing ratio of water ice

A density of particles $N$ with a particle radius $r$ is associated to a density of water ice $\chi_{H_2O}$:

$$\chi_{H_2O} = \left(\frac{4}{3} \cdot \pi \cdot r^3\right) \cdot N \cdot \frac{\rho_{ice}}{\rho_{atm}} \frac{44}{18} \quad \text{[Eq. 5]}$$

Where $\rho_{ice} = 917\ kg/m^3$ is the density of water ice and $\rho_{atm}$ is the atmospheric density. And the factor 44/18 is the ratio between the molecular weight of $CO_2$ (main compound of the Martian atmosphere), and water.

The cloud optical depth corresponding to a uniform horizontal layer covering a range of pressure $\Delta z$ is:

$$\tau = \sigma_{ext} \cdot N \cdot \Delta z = (Q_{ext} \cdot \pi \cdot r^2) \cdot N \cdot \Delta z \quad \text{[Eq. 6]}$$

Where $\sigma_{ext}$ is the Mie extinction cross section, and $Q_{ext}$ is the Mie efficiency factor.

### Relation between optical depth and number density

The total volume of ice in the air parcel ( $V_{ice}$ ) can be related to the particle radius ( $r$ ) as (neglecting the dust core and assuming for simplicity that all particles have the same radius):

$$\frac{V_{ice}}{N} = \frac{4}{3} \pi r^3 \quad \text{[Eq. 7]}$$

Introducing this in Eq. 6 we obtain:

$$\tau \propto \sqrt[3]{N \cdot V_{ice}{}^2} \quad \text{[Eq. 8]}$$

As exposed in the main text, the area where the cloud tail is present is not colder nor contains more water than the surrounding areas, the actual difference between the cloud tail and its environment is the presence of cloud particles to enable the condensation of water vapor in excess of saturation. Water ice content is lower or equal in the surrounds compared to the cloud tail (Fig. 2C). When the Aitken mode is abundant, the surrounding hazes appear because water vapor in excess of saturation condenses around nucleated Aitken dust. Consequently $V_{ice}$ is not different in the tail and in the surrounding hazes. And we can reduce Eq. 8 to:

$$\tau \propto \sqrt[3]{N} \quad \text{[Eq. 9]}$$

### Measurement of the optical depth from the shadow of the cloud

Measurement of the optical depth can be made based on images showing shadows, using the procedure previously employed by Montmessin et al.[63] and Määttänen et al.[64] This procedure compares the imaged radiance where the ground is in shadow ($R_{shadow}$ ) and where it is directly illuminated by the sun ($R_{illuminated}$ ), assuming that both areas have a similar albedo. It also takes into account the ratio between the flux of light from direct illumination by the Sun ($F_{dir}$ ) and from diffuse illumination from the sky ($F_{dif}$ ), which in Mars is mostly due to Mie scattering by dust particles in suspension. Being $\tau$ the cloud optical depth, and $\theta$ the Solar Zenith Angle, all these variables are related by this expression[63]:

$$\frac{R_{shadow}}{R_{illuminated}} = \frac{exp(\tau/cos(\theta))+F_{dif}/F_{dir}}{1+F_{dif}/F_{dir}} \quad \text{[Eq. 10]}$$

Montmessin et al.[63] estimated, using a radiative transfer model, that for an average background dust opacity of 0.2, the ratio $F_{dif}/F_{dir}$ is around 0.3.

We measured the cloud optical depth based on HRSC product HP823_0000_BL4.IMG, which was obtained on 2024-06-15 12:35 UTC. The Solar Zenith Angle of the sun on the area was $\theta$~65º.

In the HRSC image we measure that $R_{shadow}$~57 and $R_{illuminated}$~84. Solving in eq. 10, the optical depth of the AMEC in this observation was $\tau$~0.2. This would be relative to the thin hazes that could be present around the AMEC, which would reduce the value of $R_{illuminated}$. Then the actual result is more around 0.2-0.3. This should not be taken as an exact value, but as a rough estimation.

This measurement is supplementary to that reported by Fernando et al.[40] (their Fig. 9, row 1, third panel). According to their retrieval, the optical depth of the AMEC is around 0.7, and that of the hazes is around 0.1.

The difference between these two measurements could be due to the daily variations in the characteristics of the cloud.

### Data availability

This work is exclusively based on open data. Data from spacecraft observations can be found in the following repositories:

- Mars Express data is available in the ESA Planetary Science Archiva (PSA) https://psa.esa.int/
- ISRO/MCC data is publicly available under registry at https://mrbrowse.issdc.gov.in/MOMLTA/
- EXI data is publicly available under registry at https://sdc.emiratesmarsmission.ae/

Datasets produced as part of this work are available in reference(43; password of provisional repository: MF3@6iR8$9bmPSF): https://kdrive.infomaniak.com/app/share/1875470/af65aa06-1eb2-436d-920b-8f3aae53b6dc

### Code availability

This work is exclusively based on open software. The source code of the Mars PCM GCM and the Mesoscale model can be found in:

- https://svn.lmd.jussieu.fr/Planeto/trunk/LMDZ.MARS/
- https://svn.lmd.jussieu.fr/Planeto/trunk/MESOSCALE/LMD_MM_MARS/

Code produced as part of this work is available in reference(43; password of provisional repository: MF3@6iR8$9bmPSF ). https://kdrive.infomaniak.com/app/share/1875470/af65aa06-1eb2-436d-920b-8f3aae53b6dc

# Acknowledgements

We acknowledge Ehouarn Millour, Daniel Toledo, Franck Montmessin, Anna Fedorova, Miguel Ángel López Valverde, and Manuel López Puertas for their useful comments. We thank Mars Express teams (VMC in Bilbao, HRSC in Berlin, SGS in ESAC, and technical teams in ESOC) for their successful planning, acquisition and data processing. This work was supported by CNES, focused on the Mars Express and Exomars missions, and has received funding from the European Research Council (ERC) under the European Union Horizon 2020 research and innovation programme (grant agreement No 835275), project "Mars Through Time". This study benefited from the IPSL Data and Computing Center ESPRI, which is supported by CNRS, Sorbonne Université, CNES and Ecole Polytechnique

# Author contributions

Conceptualization: JHB, AM, AS, FF, Methodology: JHB, AM, AS, FF, Investigation: JHB, Visualization: JHB, Funding acquisition: AS, FF, Project administration: AS, FF, Supervision: AM, AS, FF, Writing – original draft: JHB, Writing – review & editing: JHB, AM, AS FF

# Competing interests

Authors declare that they have no competing interests.

# Materials & Correspondence

Correspondence and requests for materials should be addressed to Jorge Hernández Bernal

# Extended data

## Extended Data Table 1. List of simulations with the mesoscale model.

| **Simulations to investigate homogeneous nucleation** | | | | |
|---|---|---|---|---|
| **Simulation Name** | **ASW Surface energy** | **ASW Vapor pressure** | **Contact parameter** | **Other notes** |
| **HETERO_REF** | - | - | HTJSC | **Reference for only heterogeneous nucleation**<br>Homogeneous nucleation not activated<br>*Fig. 1B, Fig. 2B* |
| HOMO_095 | Eq. 4 | Eq. 1 | 0.95 | |
| **HOMO_HTJSC** | Eq. 4 | Eq. 1 | HTJSC | **Reference for homogeneous nucleation**<br>*Fig. 1C, Fig. 2A, Fig. 2C, Fig. 3A, Fig3B, Fig. 3F* |
| HOMO_HTTRA | Eq. 4 | Eq. 1 | HTRA | |
| NOHETERO | Eq. 4 | Eq. 1 | - | Heterogeneous nucleation deactivated<br>*Fig. 3E* |
| SIGMA_098 | 0.98 · Eq. 4 | Eq. 1 | HTJSC | Surface energy reduced by 2% |
| SIGMA_095 | 0.95 · Eq. 4 | Eq. 1 | HTJSC | Surface energy reduced by 5% |
| LOWVAP | Eq. 4 | Lowest uncertainty range of Eqs. 1 and 2 | HTJSC | *Fig. 3D* |
| LOWERVAP1 | Eq. 4 | Eq. 3 | HTJSC | Equivalent to Murray and Jensen(*5*)<br>*Fig. 3C* |
| LOWERVAP2 | Eq. 4 | Middle between Eq. 1 and Eq. 3 | HTJSC | |
| **Simulations to investigate heterogeneous nucleation on Aitken dust** | | | | |
| **Simulation Name** | **Number density ($cm^{-3}$)** | **Effective radius (nm)** | **Contact Parameter** | **Homogeneous nucleation** |
| FINEDUST_N100_R50_HTJSC | 100 | 50 | HTJSC | Deactivated |
| FINEDUST_N10_R50_HTJSC | 10 | 50 | HTJSC | Deactivated |
| FINEDUST_N1_R50_HTJSC | 1 | 50 | HTJSC | Deactivated |
| FINEDUST_N100_R50_HTTRA | 100 | 50 | HTTRA | Deactivated |
| FINEDUST_N10_R50_HTTRA | 10 | 50 | HTTRA | Deactivated |
| FINEDUST_N1_R50_HTTRA | 1 | 50 | HTTRA | Deactivated |

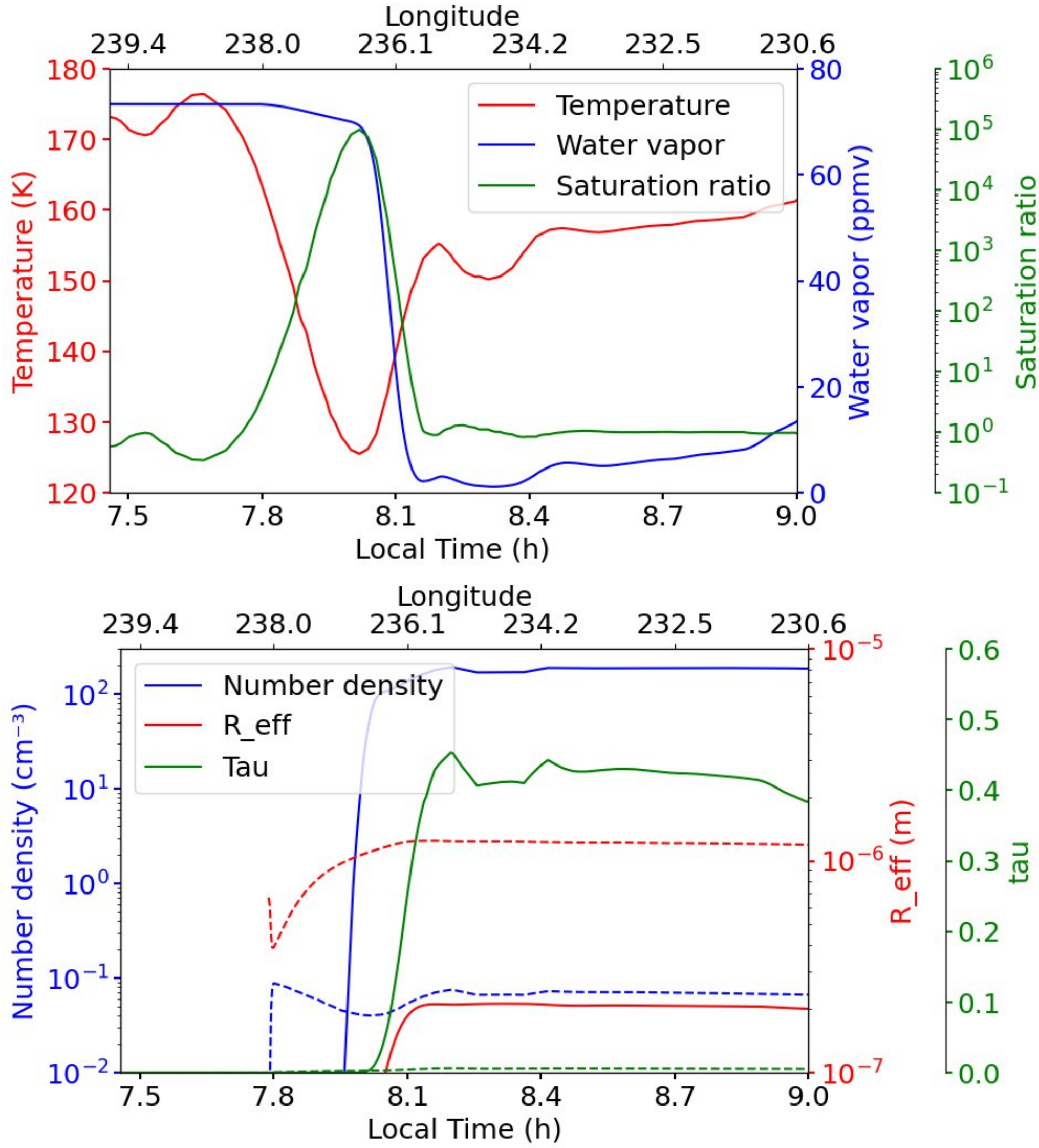

**Extended Data Figure 1. Competition between homogeneous and heterogeneous nucleation in the 0D microphysics model.** Using the default parametrizations (equivalent to HOMO_HTJSC mesoscale simulation). The curves of temperature and pressure were extracted from the HETERO_REF simulation (see Methods), for an air parcel that traverses the core cold pocket at ~8.0 a.m. The water content was set to 70ppmv and the dust number to 0.1cm$^{-3}$. Lower panel includes homogeneously (continuous curve) and heterogeneously (dashed curve) formed cloud particles. Optical depth (tau) was computed for a vertically uniform layer with 10 km of depth. Equivalent model results for the other parametrizations of the contact parameter are available in the data repository associated to this work[43], differences are negligible in this case.

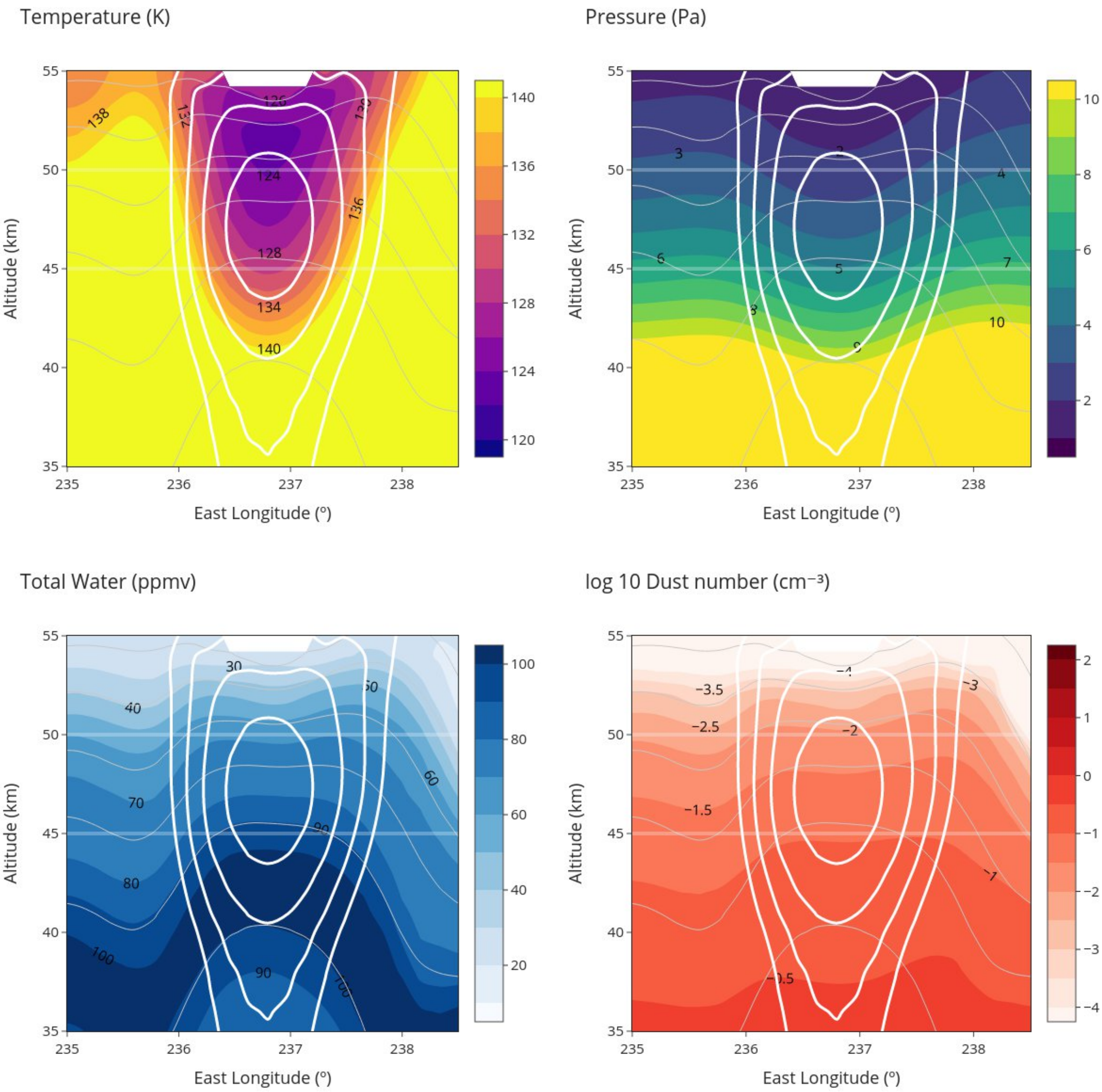

**Extended Data Figure 2. Graphs showing relevant variables for reference**. Extracted from the Reference Heterogeneous simulation (without homogeneous nucleation, at 8.1 Local Time, and latitude 8.7ºS). Thin gray curves are lines of equal potential temperature, reflecting streamlines. Thick white curves represent a temperature anomaly of -15,-20, -25 and -30K compared to the average temperature at each altitude, and they indicated the position of the core cold pocket.

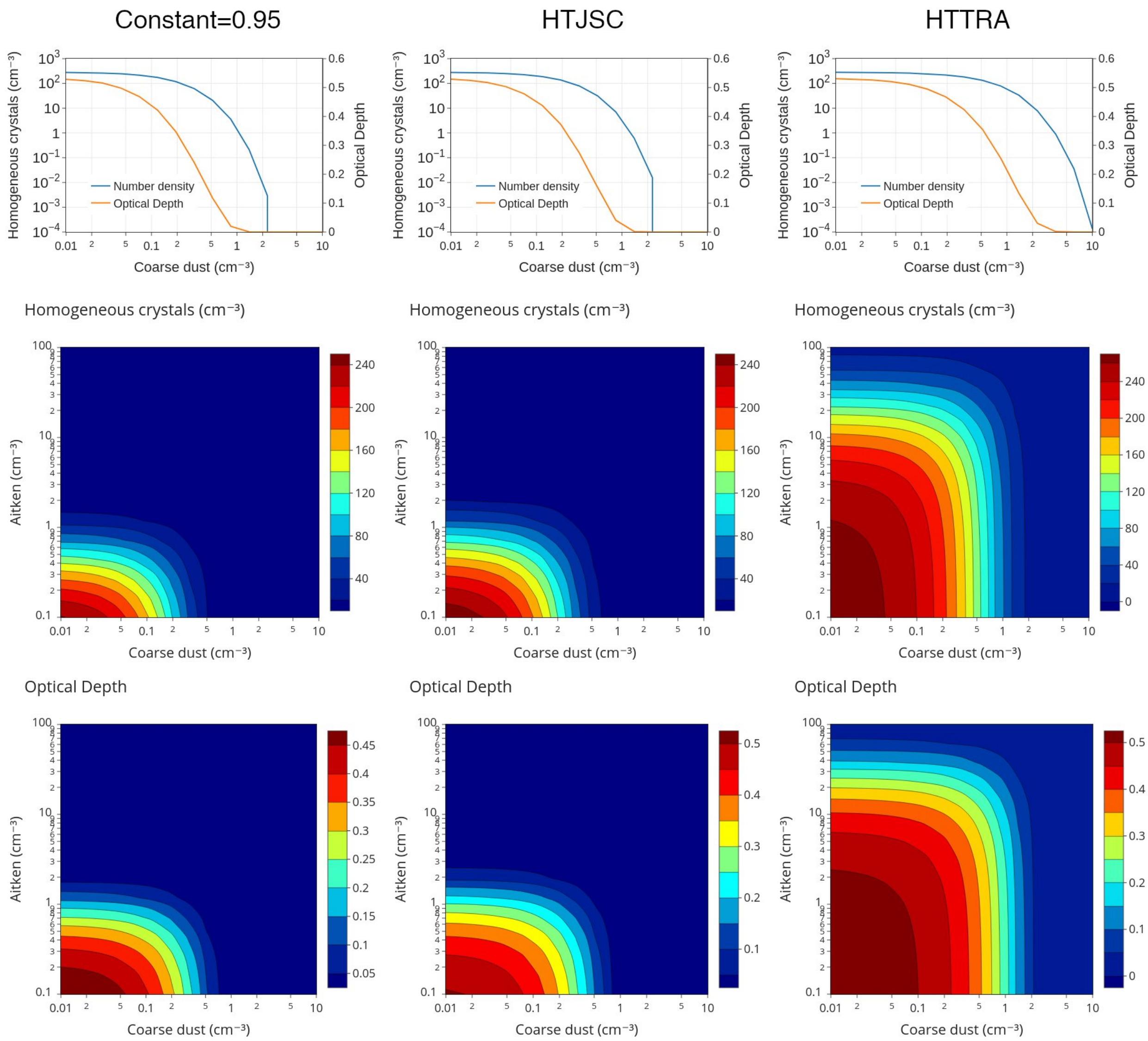

**Extended Data Figure 3. Competition between homogeneous and heterogeneous nucleation, as a function of the number density of coarse dust and Aitken dust.** Each column gives results for a different parametrization of the contact parameter. (A-C) Results for only coarse dust. (D-I) results including also Aitken dust. An optical depth of ~0.3 is achieved with ~1 $cm^{-3}$ of Aitken dust in the case of the HTJSC parametrization, and ~20 $cm^{-3}$ in the case of the HTTRA parametrization.

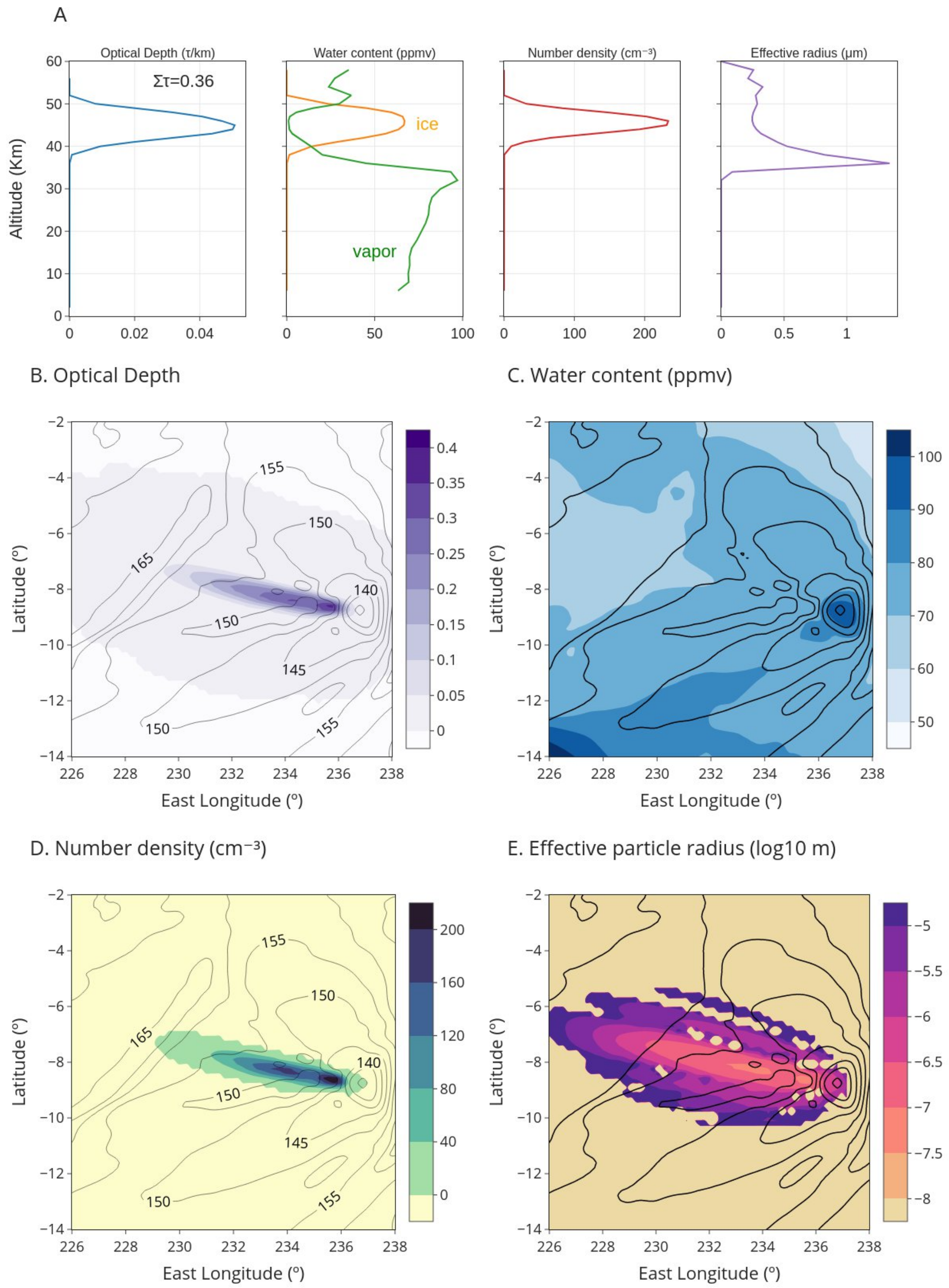

**Extended Data Figure 4. Cloud parameters.** Relation between the particle radius, number density, water availability, and optical depth. All plots are for 8.1 Local Time. (A) Vertical profile of parameters in a column close to Arsia Mons in the HOMO_HTJSC mesoscale simulation. The relatively elevated optical depth (0.36) is due to the presence of a large number density of homogeneously formed ice crystals (~200 $cm^{-3}$) with an effective particle size of ~0.25 μm. (B-E) Maps of these variables.

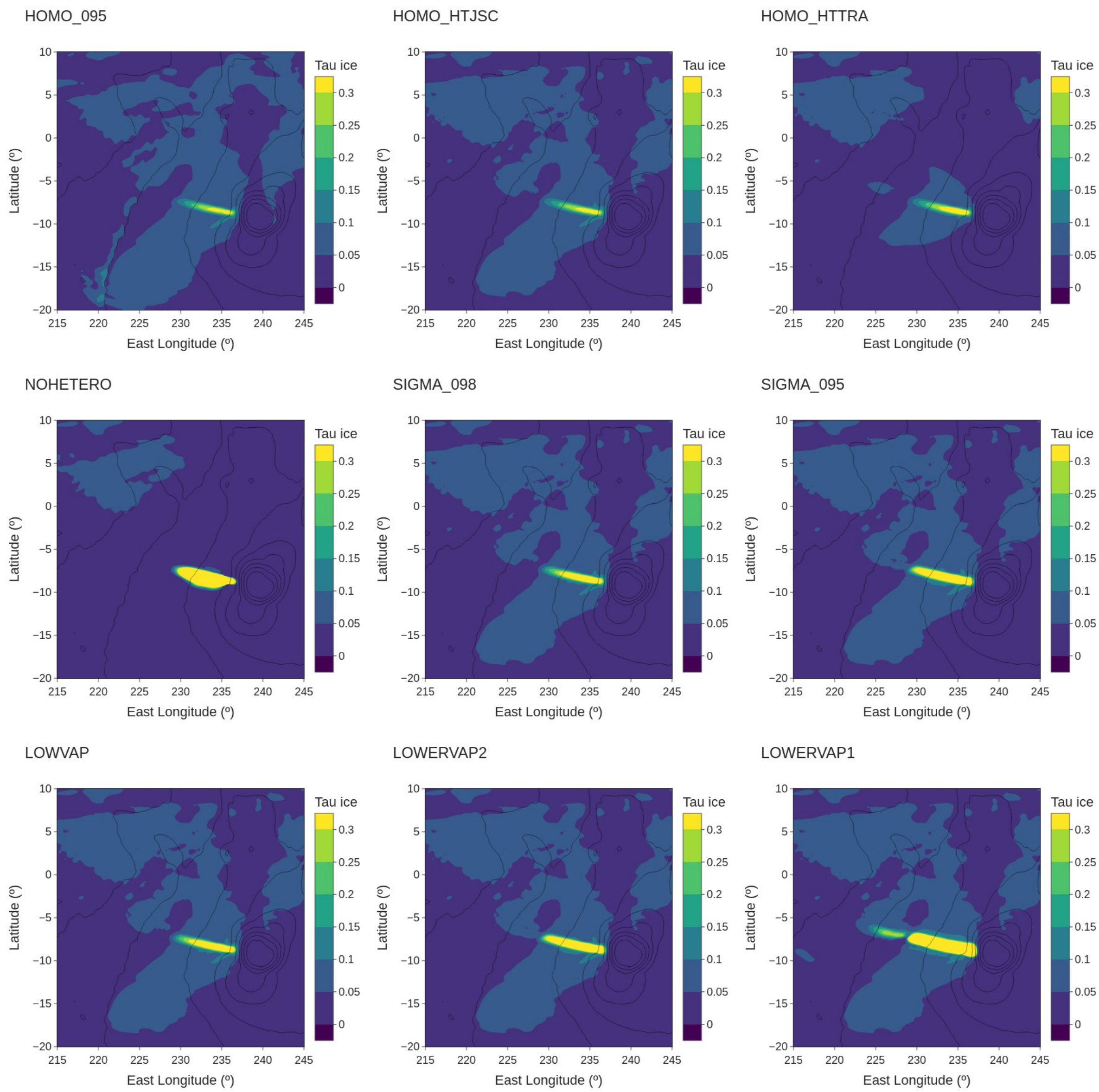

**Extended Data Figure 5. Optical depth from the mesoscale model in simulations including homogeneous nucleation.** (See table S1). All plots correspond to 8.1 a.m. Local Time. Background curves represent topography. Lowering the heterogeneous contact parameter (simulations HOMO_095, HOMO_HTJSC, HOMO_HTTRA) has a subtle impact in the optical depth of the AMEC. Reducing the surface energy by 2% and 5% (SIGMA_098, SIGMA_095) or reducing the vapor pressure (LOWVAP, LOWERVAP2, LOWERVAP1) makes the cloud brighter, wider, and a bit longer.

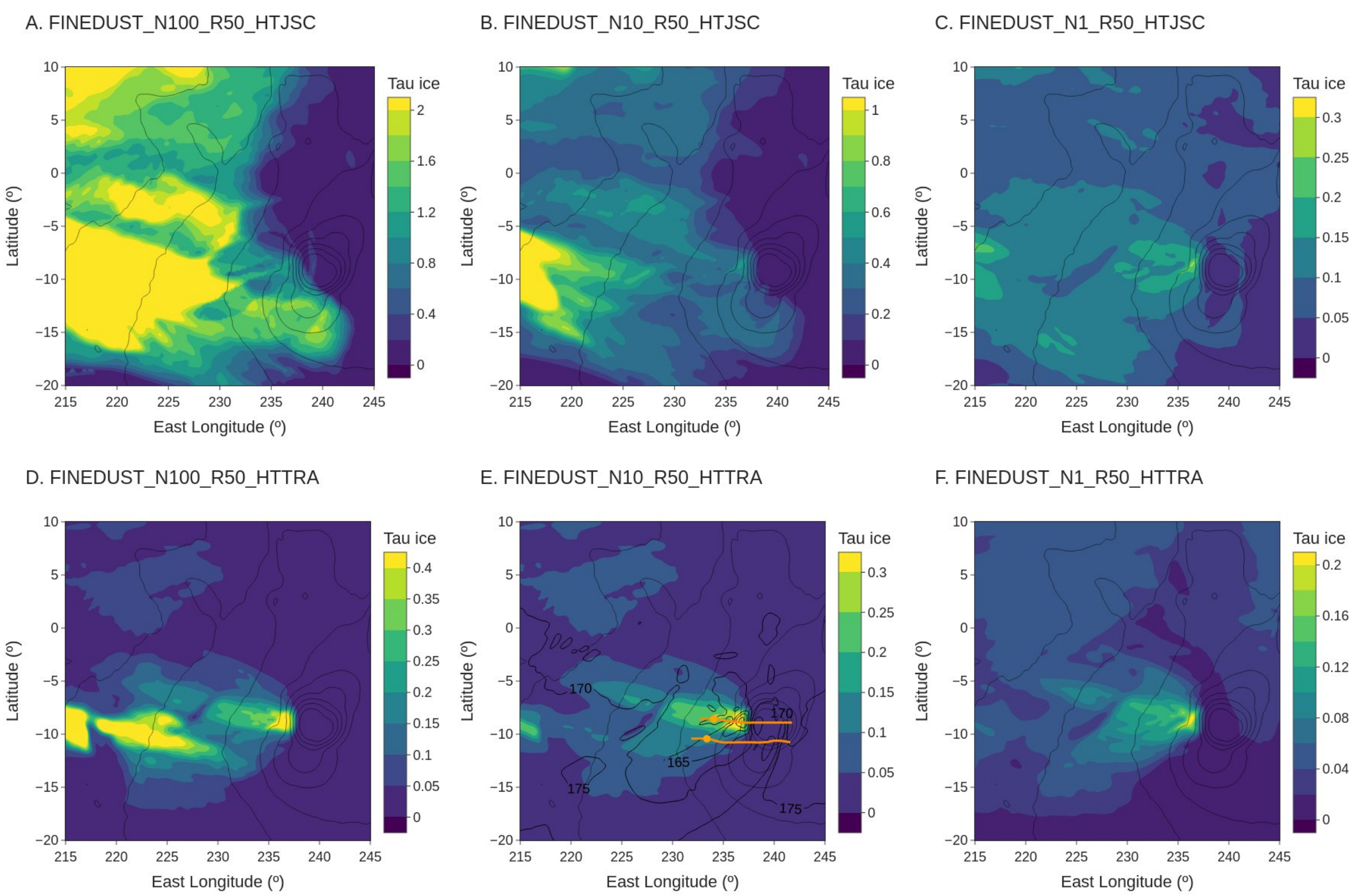

**Extended Data Figure 6. Optical depth from the mesoscale model in simulations with Aitken dust.** See table S2. All plots correspond to 8.1 a.m. Local Time. Panel C produces a cloud next to Arsia Mons whose morphology resembles cloud patterns observed instead of the AMEC during periods of enhanced dust activity (fig. S7). Panel E includes the trajectory of two lagrangian cells corresponding to simulations shown in fig. S9. In lower panels (with HTTRA contact parameter) the area downwind from the core cold pocket turns thicker than the surrounding hazes, mimicking the observed cloud tail (see main text and fig. S8).

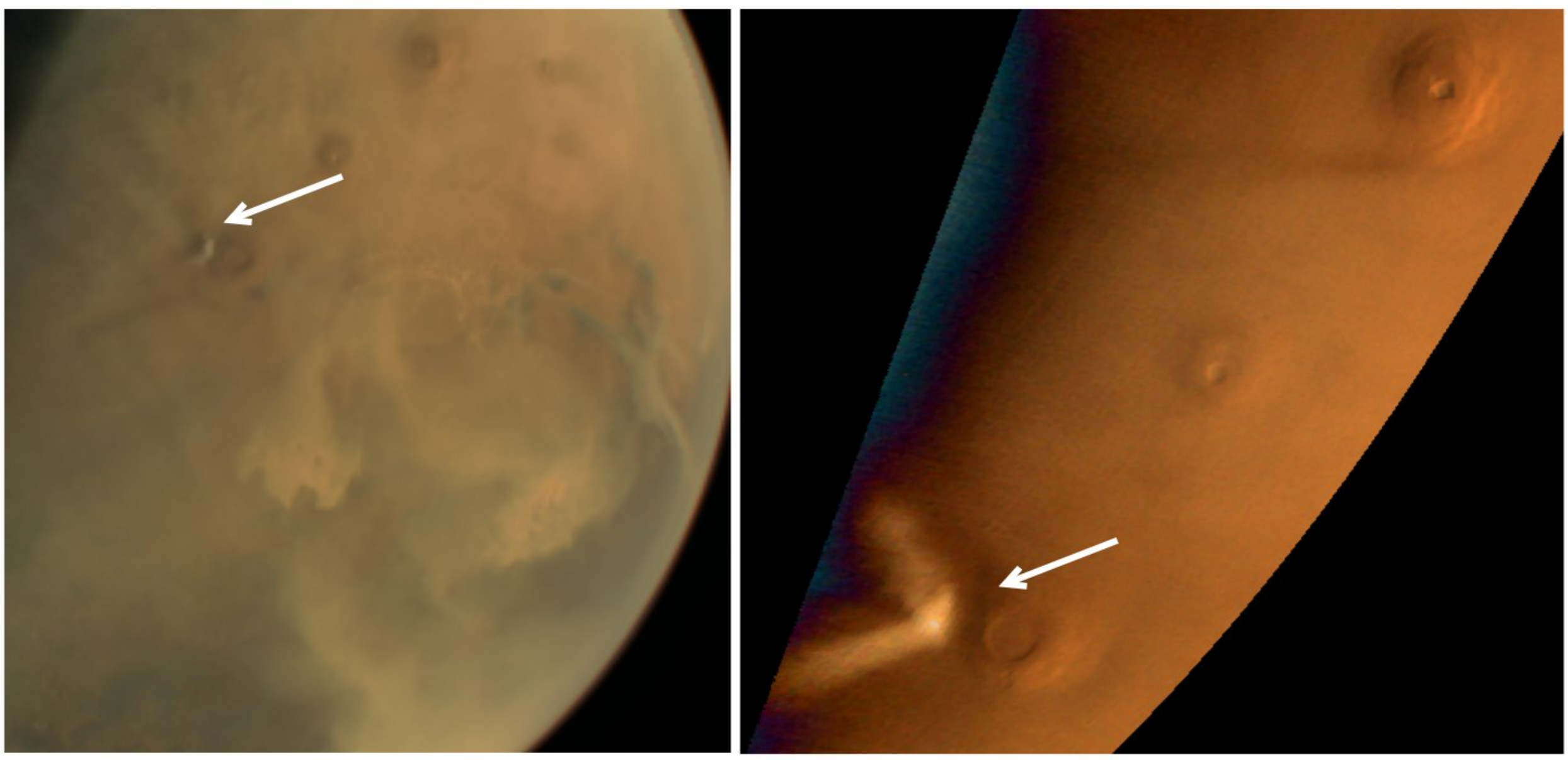

**Extended Data Figure 7. Cloud patterns on Arsia Mons during periods of the AMEC season with elevated dust activity.** Left: Image acquired by the EXI imager in orbit 275 (Ls 313, MY 36, Global dust opacity: 0.6), during the onset of a C dust event. Note that South is up in this picture, and so the area where the AMEC typically develops is to the right of Arsia Mons. Right: Image acquired by MOM/MCC on November 13 2014 (Ls 233º, MY32, Global dust opacity: 0.34), during a particularly strong A dust event that possibly delayed the starting of the AMEC season. The cloud morphology present in this image resembles the one produced by simulation FINEDUST_N1_R50_HTJSC (Fig. S6C).

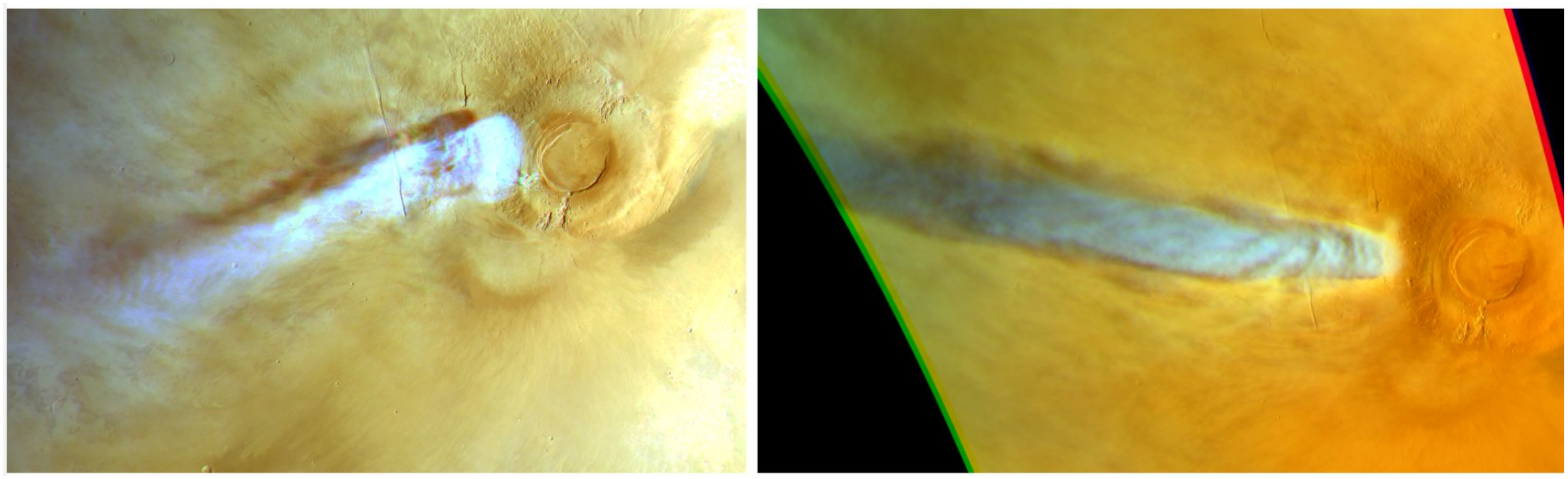

**Extended Data Figure 8. High resolution images of the AMEC acquired by MEX/HRSC** (*38,39*). (left: observation G486; right: observation P802). These pictures show that hazes around the AMEC are not comparable in brightness to the AMEC itself, and that the cloud tends to have well defined borders, features that our models only reproduce with homogeneous nucleation. See also quantitative results by Fernando et al. (*40*) (their Fig. 9, row 1, third panel).

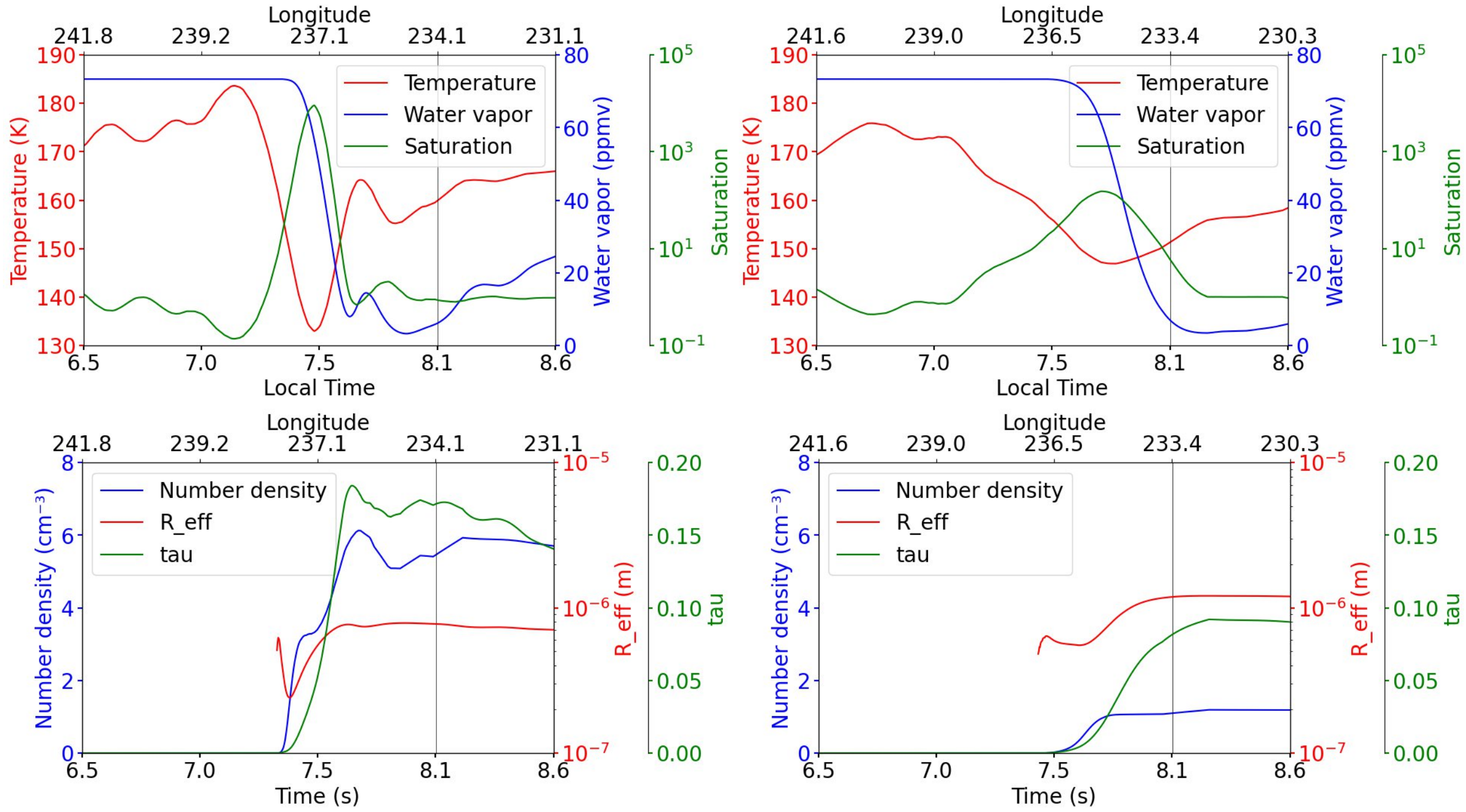

**Extended Data Figure 9. Heterogeneous nucleation on the Aitken mode in the 0D microphysics model.** These results show how heterogeneous nucleation is unable to produce the high contrast observed in optical depth. Left column corresponds to the air parcel that traverses the core cold pocket (Extended Data Figure 6E). Right column corresponds to an air parcel in the surrounding hazes (Extended Data Figure 6E). At an altitude of 40km in both cases, as this is where heterogeneous nucleation produces the higher optical depth. The extremely different temperature history experienced by both parcels is not enough to produce a very different number of cloud particles, nor a very different optical depth (as motivated by eq. 9).